\documentclass[aps,prd,nofootinbib,showkeys,preprint,floatfix]{revtex4}

\usepackage{amssymb}
\usepackage{amsmath}
\usepackage{amsfonts}
\usepackage{graphicx}
\usepackage{color}
\usepackage{xspace}
\usepackage{ulem}
\usepackage{lscape}
\usepackage{ wasysym }

\usepackage{multirow,rotating}
\usepackage{dsfont}

\def\gsim{\raise0.3ex\hbox{$\;>$\kern-0.75em\raise-1.1ex\hbox{$\sim\;$}}}
\def\lsim{\raise0.3ex\hbox{$\;<$\kern-0.75em\raise-1.1ex\hbox{$\sim\;$}}}

\def\znbb{0\nu\beta\beta}
\def\meff{\langle m_{\nu} \rangle}

\newcommand{\ba}[1]{\begin{eqnarray} \label{(#1)}}
\newcommand{\ea}{\end{eqnarray}}

\newcommand{\AddrAHEP}{
  {\it AHEP Group, Instituto de F\'{\i}sica Corpuscular --
    CSIC./Universitat de Val{\`e}ncia \\
    Edificio de Institutos de Paterna, Apartado 22085,
  E--46071 Val{\`e}ncia, Spain}}

\newcommand{\AddrUFSM}{
Universidad T\'ecnica Federico Santa Mar\'\i a, \\ 
Centro-Cient\'\i fico-Tecnol\'{o}gico de Valpara\'\i so, \\ 
Casilla 110-V, Valpara\'\i so,  Chile}

\def\gsim{\raise0.3ex\hbox{$\;>$\kern-0.75em\raise-1.1ex\hbox{$\sim\;$}}}
\def\lsim{\raise0.3ex\hbox{$\;<$\kern-0.75em\raise-1.1ex\hbox{$\sim\;$}}}

\begin{document}

\preprint{IFIC/15-80}  

\title{QCD running in neutrinoless double beta decay: Short-range mechanisms}

\author{M. Gonz\'alez} \email{marcela.gonzalezp@usm.cl}\affiliation{\AddrUFSM}
\author{M. Hirsch} \email{mahirsch@ific.uv.es}\affiliation{\AddrAHEP}
\author{S.G. Kovalenko}\email{sergey.kovalenko@usm.cl}\affiliation{\AddrUFSM}



\begin{abstract}
The decay rate of neutrinoless double beta ($0\nu\beta\beta$) decay
contains terms from heavy particle exchange, which lead to dimension-9
($d=9$) six fermion operators at low energies. Limits on the
coefficients of these operators have been derived previously
neglecting the running of the operators between the high-scale, where
they are generated, and the energy scale of $\znbb$-decay, where
they are measured.  Here we calculate the leading order QCD
corrections to all possible $d=9$ operators contributing to the
$0\nu\beta \beta$ amplitude and use RGE running to calculate 1-loop
improved limits. Numerically, QCD running dramatically changes some limits by factors of
the order of or larger than typical uncertainties in nuclear matrix
element calculations.  For some specific cases, operator mixing in the
running changes limits even by up to three orders of magnitude.  Our
results can be straightforwardly combined with new experimental limits
or improved nuclear matrix element calculations to re-derive updated
limits on all short-range contributions to $0\nu\beta\beta$ decay.
\end{abstract}

\maketitle


%
\section{Introduction}
\label{Sec:Int}

Non-observation of neutrinoless double beta ($\znbb$) decay constrains lepton
number violating extensions of the Standard Model (SM). Usually lower
limits on $\znbb$ decay half-lives are interpreted as upper limits on
the effective Majorana neutrino mass, $\meff = \sum_j m_j U_{ej}^2$,
but many models generating a non-zero $\znbb$ decay amplitude not
directly proportional to $\meff$ have been discussed in the
literature, for recent reviews on $\znbb$ decay see for example
\cite{Avignone:2007fu,Deppisch:2012nb}.

One can classify the different contributions to the general $\znbb$
decay rate either as long-range \cite{Pas:1999fc} or as short-range
\cite{Pas:2000vn} contributions.  The long-range part of the amplitude
describes the exchange of a light neutrino between two point-like
vertices. If both vertices are the SM charged current vertices, the
resulting diagram corresponds to the well-known mass mechanism, but
other long range contributions, not directly proportional to $\meff$,
do exist in many models, like for example R-parity violating SUSY
\cite{Babu:1995vh,Hirsch:1995cg,Pas:1998nn} or leptoquark models~\cite{Hirsch:1996ye}.

The short-range part of the $\znbb$ amplitude is due to ``heavy''
particle exchange.
\footnote{Any particle with mass larger than the typical Fermi
  momentum of the nucleons, i.e. ${\cal O}(0.1)$ GeV, can be
  considered ``heavy'' in $\znbb$ decay.  All exotic fermions
  contributing to the short-range amplitude, except possibly sterile
  neutrinos, are expected to have masses larger than ${\cal O}(100
  \hskip1mm{\rm GeV})$.}  
After integrating them out the amplitude can be represented as (the
nuclear matrix element of) a true dimension-9 ($d=9$) quark-level
effective operator, which can be schematically written as:
\begin{equation}
\mathcal{O}_{d=9}
\propto 
\frac{1}{\Lambda_{\rm LNV}^5} \bar u \bar u  \, d d \, \bar e \bar e \ .
\label{eq:dim9}
\end{equation}
The general $SU(3)_c\times SU(2)_L \times U(1)_Y$ invariant
decomposition of this $d=9$ operator has been discussed in
\cite{Bonnet:2012kh}. The tables given in \cite{Bonnet:2012kh} can be
understood as a summary of all \mbox{(proto-)} high-energy scale models, which contribute to
$\znbb$ decay at tree-level via heavy particle exchange. Once all
possible ultraviolet (UV) completions of Eq.~(\ref{eq:dim9}) have been specified, one
can then use the results of \cite{Pas:2000vn} to derive general limits
on all possible models contributing to $\znbb$ decay.

Given current experimental lower limits on half-lives of $\znbb$
decay, of the order of (few) $10^{25}$ ys for $^{76}$Ge
\cite{Agostini:2013mzu} and $^{136}$Xe
\cite{Albert:2014awa,Shimizu:2014xxx,Gando:2012zm}, the energy scale,
$\Lambda_{\rm LNV}$, at which the effective interactions (\ref{eq:dim9}) are generated, is
expected to be of the order of roughly ${\cal O}$(TeV). On the other hand, $\znbb$ decay
is a low-energy process with the typical momentum 
scale given by the Fermi momentum of nucleons, $p_F\sim 100$ MeV. This rather large
mismatch in scales implies that the running of the operators under the renormalization may be
quite important numerically. This observation forms the basic
motivation for the current paper.

The Operator Product Expansion (OPE) and the renormalization group equation (RGE) have become the standard tool \cite{Buchalla:1995vs,Buras:1998raa} in electro-weak precision
physics.  Here, we extend this formalism to  $\znbb$-decay.
We shall specify {\em all possible $d=9$ operators}  contributing to $\znbb$ decay
and calculate their QCD leading order RGE running.
Color mismatched operators, which appear in this calculation, lead to 
operator mixing. Since different operators in $\znbb$ decay can 
have vastly different nuclear matrix elements, this effect in 
some case leads to a rather drastic change in the derived limits. 
It is therefore important to take these QCD corrections into 
account in the calculation of limits on short-range operators. 

We note, that our paper is not the first to consider QCD corrections.
In \cite{Mahajan:2014twa} the author 
observed that
the color mismatch generated by the QCD corrections is expected to be important in the case of the scalar-pseudoscalar quark operators ($\mathcal{O}_{1}$ in the notations of Eq.~(\ref{eq:OperBasis-1})).
Reference
\cite{Peng:2015haa} 
treats in details the running and mixing of the scalar-pseudoscalar and tensor operators 
($\mathcal{O}_{1,2}^{RR}$ in the notations of Eqs. (\ref{eq:OperBasis-1})-(\ref{eq:OperBasis-2})).
Our current paper, however, is the first one to give the leading order QCD corrections to the complete set of the short-range $d=9$ $\znbb$-operators covering the low-energy limits of any possible underlying high-energy scale model.

The rest of this paper is organized as follows. In the next
section we remind the most important definitions for
operators, currents and the $\znbb$-decay half-life given in
\cite{Pas:2000vn}, before summarizing in section \ref{Sec:Decomp}
how to connect low-energy $\znbb$-decay with the possible
ultra-violet completions (``models'') of the $d=9$ operators
\cite{Bonnet:2012kh}.  Section \ref{sec:QCD}
describes the formalism of effective theories based on the operator product expansion and the renormalization group, which we use in the analysis of $\znbb$-decay. 
Section \ref{Sec:Num} contains the central result of the present paper: the leading order {\it QCD corrected 
$\znbb$-decay half-life  formula} (\ref{eq:T12-CMW}). Here we also 
discuss our numerical results, before closing with a short summary
in section \ref{Sec:Cncl}.


\section{Low-energy Effective Lagrangian and $\znbb$-decay Half-life}
\label{Sec:Ops}

From the low-energy point of view, adequate for the energy scale
$\mu\sim 100$~MeV of $0\nu\beta\beta$-decay, the short range (SR)
part of the decay amplitude can be derived from the generic effective
Lagrangian \cite{Pas:2000vn}
\footnote{In \cite{Pas:2000vn} the coefficients in
  Eq. (\ref{eq:LagGen}) where denoted as $\epsilon_{i}^{XY}$.}
\begin{eqnarray}\label{eq:LagGen}
{\cal L}^{0\nu\beta\beta}_{\rm eff} = \frac{G_F^2}{2 m_p} \,
              \sum_{i, XY} C_{i}^{XY}(\mu)\cdot \mathcal{O}^{XY}_{i}(\mu),
\end{eqnarray}
with the $d=9$ operator basis containing the following complete set of  Fierz non-equivalent operators, classified 
by their Lorentz structure:
\begin{eqnarray}
\label{eq:OperBasis-1}
\mathcal{O}^{XY}_{1}&=& 4 ({\bar u}P_{X}d) ({\bar u}P_{Y}d) \ j,\\
\label{eq:OperBasis-2}
\mathcal{O}^{XX}_{2}&=& 4 ({\bar u}\sigma^{\mu\nu}P_{X}d)
                         ({\bar u}\sigma_{\mu\nu}P_{X}d) \ j,\\
\label{eq:OperBasis-3}
\mathcal{O}^{XY}_{3}&=& 4 ({\bar u}\gamma^{\mu}P_{X}d) 
                        ({\bar u}\gamma_{\mu}P_{Y}d) \  j,\\
\label{eq:OperBasis-4}
\mathcal{O}^{XY}_{4}&=& 4 ({\bar u}\gamma^{\mu}P_{X}d) 
                         ({\bar u}\sigma_{\mu\nu}P_{Y}d) \ j^{\nu},\\
\label{eq:OperBasis-5}
\mathcal{O}^{XY}_{5}&=& 4 ({\bar u}\gamma^{\mu}P_{X}d) ({\bar u}P_{Y}d) \ j_{\mu}
\end{eqnarray}
with $X,Y = L,R$ and the leptonic currents are 
\begin{eqnarray}\label{eq:Curr}
%
j = {\bar e}(1\pm \gamma_{5})e^c \, , \quad j_{\mu} = {\bar e}\gamma_{\mu}\gamma_{5} e^c .
\end{eqnarray}
Let us stress that the so-called color mismatched operators appearing in low-energy limit of high-scale models as well as due to the QCD corrections,
as explained in sec.~\ref{sec:QCD},  can all be expressed in terms of the color-singlet  operator basis 
(\ref{eq:OperBasis-1})-(\ref{eq:OperBasis-5}). An example of a color mismatched operator and its expression in terms of this operator basis is given in Eqs.~(\ref{eq:step2}),~(\ref{eq:step3}). 

The following notes on the effective Lagrangian (\ref{eq:LagGen}) and the operator basis (\ref{eq:OperBasis-1})-(\ref{eq:OperBasis-5}) are also in order. The leptonic currents $\bar{e} \gamma^{\mu} e^{c}$, $\bar{e}
\sigma^{\mu\nu} e^{c}$ and $\bar{e} \sigma^{\mu\nu}\gamma_{5} e^{c}$
vanish identically. That is why they do not appear in Eqs.~(\ref{eq:OperBasis-1})-(\ref{eq:Curr}). For the current $j = \bar{e}(1\pm \gamma_{5})e^{c}$  in Eq.~(\ref{eq:Curr}) we used notation without distinguishing the relative sign. This is because the $\znbb$-decay half-life, given below in Eq.~(\ref{eq:T12}), does not depend on it.   
Note
further that the factor $\frac{G_F^2}{2 m_p}$ in Eq.~(\ref{eq:LagGen})
has been chosen to make the coefficients $C_{i}$ dimensionless
quantities and we have introduced a factor of 
$4$ in Eqs.~(\ref{eq:OperBasis-1})-(\ref{eq:OperBasis-5}), such that the numerical
values of $C_{i}$ can be directly compared with the numbers given
in the original paper \cite{Pas:2000vn}. Finally, all the
operators (\ref{eq:OperBasis-1})-(\ref{eq:OperBasis-5}) can have superscripts $XY$ with the exception
of $\mathcal{O}^{XX}_{2}$, for which
$\mathcal{O}^{LR}_{2}=\mathcal{O}^{RL}_{2}\equiv 0$.

Eq.~(\ref{eq:LagGen}) is nothing but the most general parametrization
of the effective Lagrangian in terms of the quark-lepton operators of the lowest dimension, $d=9$,
which can contribute to the $0\nu\beta\beta$-decay amplitude at tree
level.  No particular physics underlying the Lagrangian
(\ref{eq:LagGen}) is implied at this stage.  Note that the Lagrangian
(\ref{eq:LagGen}) is tied to the typical energy scale $\mu$ of the process in question.
For $0\nu\beta\beta$-decay it is of the order of the Fermi momentum
of nucleons and quarks in $0\nu\beta\beta$-decaying nucleus, \mbox{$\mu\sim p_{F}
\sim$ $100$ MeV.}  The Lagrangian (\ref{eq:LagGen}) can be applied to
a processes with any hadronic states: quarks, mesons, nucleons,
other baryons and nuclei. The corresponding amplitude is determined by
the hadronic matrix elements of  
the operators $\mathcal{O}_{i}$ of in Eqs. (\ref{eq:OperBasis-1})-(\ref{eq:OperBasis-5}). The coefficients $C_{i}$ are independent of the low-energy scale
non-perturbative hadronic dynamics. This is the well-recognizable
feature of the Operator Product Expansion (OPE), representing
interactions of some high-scale renormalizable model in the form of
Eq.~(\ref{eq:LagGen}) below a certain scale $\mu$. The coefficients
$C_{i}$ are known as Wilson coefficients, depending on the parameters
of a high-scale model.

Applying the standard nuclear theory methods \cite{Doi:1985dx}, one finds for the $\znbb$ half-life:
\begin{eqnarray}\label{eq:T12} 
\Big[ T^{\znbb}_{1/2}\Big]^{-1} = G_1 
\left|\sum_{i=1}^{3} C_{i}(\mu_{0}) {\cal M}_{i}\right|^2 +  G_2 
\left|\sum_{i=4}^{5} C_{i}(\mu_{0}) {\cal M}_{i}\right|^2
\end{eqnarray}
Here, $G_1=G_{01}$ and $G_2=(m_eR)^2 G_{09}/8$ are phase space factors
in the convention of \cite{Doi:1985dx}. Their numerical values for various isotopes can be found 
in Ref.~\cite{Deppisch:2012nb}. The quantities \mbox{${\cal M}_i = \langle
A_{f}| \mathcal{O}^{\rm h}_{i}| A_{i}\rangle$} are the nuclear matrix
elements defined in Ref.~\cite{Pas:2000vn}. In the above equation the
summation over the coefficients corresponding to the operators ${\cal
  O}^{XY}_{i}$ with different chiralities $X,Y = L,R$ is implied.  The
Wilson coefficients $C_{i}(\mu_{0})$ should be taken close to the
typical $\znbb$-energy scale. In our analysis we choose $\mu_{0} =
1$GeV.  In Eq. (\ref{eq:T12}) we have not included interference terms,
since none of the
high-scale models listed in \cite{Bonnet:2012kh} mixes the
coefficients $C_{1,2,3}$ with $C_{4,5}$.

Numerical values for the nuclear matrix elements ${\cal M}_{i}$, based
on the pn-QRPA approach of \cite{Muto:1989cd}, can be found for
$^{76}Ge$ in \cite{Pas:2000vn}, for other isotopes of interest see
\cite{Deppisch:2012nb}. It is, however, well-known that nuclear matrix
elements for $\znbb$-decay have quite large numerical
uncertainties. Recent publications calculating matrix elements for
heavy neutrino exchange, i.e. matrix elements for the short-range part
of the amplitude corresponding in our notation to the term
$C^{LL}_{3}$, give numerical values which are larger than those of
\cite{Deppisch:2012nb} by typically 50 \% (40 \%) in the QRPA
calculation by the T\"ubingen group \cite{Faessler:2014kka}
(Jyv\"askyl\"a group \cite{Hyvarinen:2015bda}). Shell model
calculations for light neutrino exchange, on the other hand, seem to
give matrix elements which are up to a factor of two smaller than
those of QRPA \cite{Menendez:2011zza}.  Similar factors are found for
heavy neutrino exchange in the shell model calculation of
\cite{Blennow:2010th}. However, a recent shell model calculation for
$^{76}$Ge gives matrix elements for light neutrino exchange 
\cite{Sen'kov:2014jfa} only 15-40~\% smaller than than those of
\cite{Muto:1989cd}. While these variations in numerical results do
probably not cover the error bar in the calculation of nuclear matrix
elements completely, from these numbers one may estimate that
currently matrix elements for the short-range part have uncertainties
of roughly a factor of 2 or so.

We note, however, that while we do use the numerical values of
\cite{Deppisch:2012nb} for the derivation of new limits, all our
calculations are presented in such a way that the running of the
operators is separated completely from the nuclear structure part of
the calculation. Thus, our coefficients can be combined with any new
nuclear matrix element calculations, should they become available, to
extract updated limits. For the time being the numerical values for  nuclear matrix 
of the whole set of the basis operators in Eqs. (\ref{eq:OperBasis-1})-(\ref{eq:OperBasis-5}) are not available 
in the literature in the approaches other than that of Refs.~\cite{Pas:2000vn,Deppisch:2012nb}. However, we have recently learned~\cite{Simkovic:2015NME} that the corresponding results within the QRPA approach of the
T\"ubingen group will be published soon.

\section{Link to High-Scale Models}
\label{Sec:Decomp}

As already mentioned above, Eq.~(\ref{eq:LagGen}) is a general
parametrization of all the possible $d=9$ contact interactions contributing
to $\znbb$-decay amplitude at tree level, without referring to any
underlying physics. The latter is typically thought to be represented
by renormalizable models with heavy degrees of freedom which decouple
from the light sector at certain energy scale (much) larger than the
characteristic scale of $\znbb$-decay.  In the literature one can find
two approaches connecting the effective Lagrangian (\ref{eq:LagGen})
to such high-energy models. We will discuss them briefly.

\begin{figure}[t]
\hskip-10mm\includegraphics[width=0.5\linewidth]{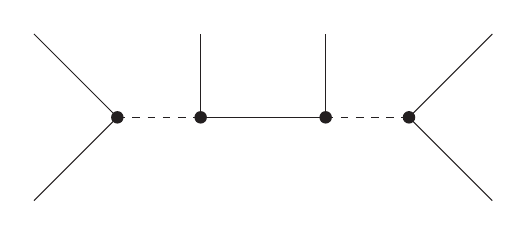}
\hskip10mm\includegraphics[width=0.4\linewidth]{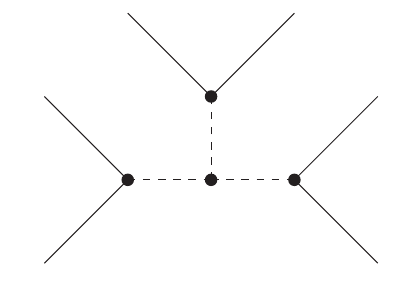}
\caption{Tree-level topologies contributing to 
the $\znbb$-decay rate. To the left T-I, scalar-fermion-scalar 
exchange; to the right T-II, scalar-scalar-scalar diagrams. 
Scalars could also be replaced by vectors.}
\label{fig:topos}
\end{figure}

Historically, the first approach was the top-down approach: Starting
from a concrete high-scale model and integrating out heavy degrees of
freedom of a mass $M_{h}$ at energy-scales $\mu < M_{h}$. Then, there
appear effective non-renormalizable interactions of the light fields
in the form of an expansion in the inverse powers of $M_{h}$, which is
the operator product expansion. The interactions (\ref{eq:LagGen}) are
then the leading $d=9$ terms of this expansion.  The well-known and
simplest example of such a model is the SM, extended by a heavy
Majorana neutrino $N$ with the mass $M_{N}\gg \mu\sim
p_{F}\sim$ $100$ MeV. The relevant Lagrangian term is
\begin{eqnarray}\label{eq:MN}
{\cal L}_{\rm SMN} = 
\frac{g_{2}}{\sqrt{2}}\, \overline{e_{L}}\, \gamma^{\mu} U_{eN} N\cdot  W^{-}_{\mu}
\end{eqnarray}
where $g_{2}$ is the $SU(2)_{L}$ gauge coupling constant and $U_{eN}$
describes the mixing of this heavy state with the ordinary electron
neutrino. The tree-level diagram contributing to the $\znbb$ amplitude
is shown in Fig. (\ref{fig:topos}) on the left with $W^{\pm}$ denoted
by dashed lines.  At momenta $p$ of the external legs below both
$M_{W}$ and $M_{N}$ one can expand the corresponding propagators in
$p^{2}/M^{2}_{i}$ with $i = W, N$. The leading term is
\begin{eqnarray}\label{eq: example-N}
{\cal L}_{SMN}^{\rm eff} = -8 \frac{G_{F}}{\sqrt{2}} \frac{U_{eN}}{M_{N}}\, 
\overline{u_{L}}\, \gamma_{\mu} d_{L}\cdot  
\overline{u_{L}} \, \gamma_{\mu} d_{L}\cdot 
\bar{e}P_R e^{c}  .
\end{eqnarray}
In the path integral approach the described procedure is equivalent to
integrating out the $W$ and $N$ fields, which consists of neglecting
their kinetic terms, justified at energies below their masses, and the
subsequent Gaussian integration over $W$ and $N$ variables. (For a
pedagogical review see Refs.~\cite{Buchalla:1995vs, Buras:1998raa}).
In the literature a great host of high-scale models have been linked
to the form of the Lagrangian (\ref{eq:LagGen}) in this way. The key
point here is that there are at least two orders of magnitude of
hierarchy between the scale where the new degrees of freedom are
integrated out and the scale of $\znbb$-decay, which the
parameterization of Eq.~(\ref{eq:LagGen}) is tied to.  As has been
pointed out for the first time in Ref. \cite{Mahajan:2014twa} in the
presence of QCD loop corrections such a scale hierarchy has a
significant impact on the relation of the parameters of high-scale
models and the parameters $C_{i}$ extracted from the measurements of
$\znbb$-decay half-life on the basis of Eq.~(\ref{eq:T12}).

Recently, a bottom-up approach to ``deconstructing'' $\znbb$-decay has
been proposed in Ref.~\cite{Bonnet:2012kh}. This 
approach 
surveys in a generic way all possible
renormalizable \mbox{$SU(3)_c\times SU(2)_L \times U(1)_Y$} invariant
interactions leading in the low energy limit to the effective
operators in Eqs.~(\ref{eq:LagGen})-(\ref{eq:OperBasis-5}). As demonstrated in \cite{Bonnet:2012kh},
there are only two tree-level topologies for the renormalizable decompositions of these
operators.
These are shown in
Fig.~\ref{fig:topos} and denoted T-I and T-II. The six outside lines
stand for any of ${\bar u}$, $d$ or ${\bar e}$. Dashed lines are for
bosons (either scalars or vectors), the solid (inner) line in T-I is
for some exotic (i.e.  non-standard model) fermion.
The task of defining all possible ultraviolet completions
(``models'') contributing to the $\znbb$-decay rate (at tree-level),
then reduces the problem to finding all SM-invariant fermion bilinears
involving the quarks and leptons (plus all bilinears involving one SM
fermion and one exotic fermion in case of T-I) of Eq.~(\ref{eq:dim9})
and, after integrating out all heavy (i.e. beyond SM) particles,
rewrite the resulting expressions into the basis operators of
Eq.~(\ref{eq:LagGen}). 

To make contact with some of the known in the literature mechanisms of  $\znbb$-decay and to the general treatment 
of all the possible short range mechanisms \cite{Bonnet:2012kh}, we discuss here a
simple example model based on the decomposition T-I-1-i in the notations of Ref.~\cite{Bonnet:2012kh}. This corresponds to
the external legs in Fig.~\ref{fig:topos}~(left) grouped as $({\bar u_L}d_R)
({\bar e_L})({\bar e_L})({\bar u_L}d_R)$, so that the  fields  within  the same parenthesis meet in the same vertex. With this choice, the two
scalars $S_1=S_2$ are fixed to be either $S_{1,2,1/2}$ or
$S_{8,2,1/2}$. For the former the intermediate fermion is either
$\psi_{1,1,0}$ or $\psi_{1,3,0}$, while for the latter it is either 
$\psi_{8,1,0}$ or $\psi_{8,3,0}$. 
This model partially covers the well known case of the R-parity violating SUSY mechanism \cite{Hirsch:1996ye}
identifying
$S_{1,2,1/2} \equiv \tilde{L}$ and $\psi_{1,1,0}\equiv \chi_{0}$ with $\tilde{L} = (\tilde{\nu},\tilde{e})_{L}$  and 
$\chi_{0}$ being the slepton electroweak doublet, scalar superpartner of  the left-handed lepton doublet, and the neutralino, respectively. 
Conventionally, the subscripts denote the field assignment to certain representation of the SM gauge group 
$SU(3)_c\times SU(2)_L \times U(1)_Y$ in the format \mbox{(dimension, dimension, $U(1)_{Y}$-charge)}.
%
The model interactions appearing in  the
diagram T-I in Fig.~\ref{fig:topos}~(left) are then given by:
\begin{eqnarray}\label{eq:LagExa} 
{\cal L}_Y &=& Y_{Qd(1)}({\bar Q}d_R)S_{1,2,1/2} + 
             Y_{Qd(8)}\left({\bar Q}\frac{\lambda^A}{2} d_{R}\right) S^{A}_{8,2,1/2}
\\ \nonumber
        & + & Y_{e\psi(1)}({\bar e_L}\psi_{1,X,0})S_{1,2,1/2}^{\dagger} 
          +  Y_{e^c\psi(1)}({\bar e_L^c}\psi_{1,X,0})S_{1,2,1/2}
\\ \nonumber
 & +&  Y_{e\psi(8)}({\bar e_L}\psi^{A}_{8,X,0}) S_{8,2,1/2}^{A\dagger} 
   +  Y_{e^c\psi(8)}({\bar e_L^c}\psi^{A}_{8,X,0})S^{A}_{8,2,1/2}.
%
\end{eqnarray}
Here, $(\lambda^A)$ are the Gell-Mann matrices, and $Y$ some 
unknown Yukawa couplings. Eq.~(\ref{eq:LagExa}), together with the 
Majorana propagator for $\psi_{C,X,0}$ and after integrating out 
heavy particles, gives an effective Lagrangian, which for the 
color octet case, $S_{8,2,1/2}$,  reads
\begin{eqnarray}\label{eq:LagExaEff}
{\cal L}_{eff} = 
\frac{Y_{Qd(8)}^2Y_{e\psi(8)} Y_{e^c\psi(8)}}{m^4_{S_{8,2,1/2}}m_{\psi_{8,X,0}}}
(\lambda^A/2)^{b}_{a}(\lambda^A/2)^{d}_{c}({\bar Q}^{a}d_{R,b})
({\bar Q}^{c}d_{R,d})({\bar e}P_R e^c).
\end{eqnarray}
The Lagrangian for the color singlet case is identical to Eq.~(\ref{eq:LagExaEff}) after some obvious replacements,
i.e. $(\lambda^A/2)^{i}_{j}\rightarrow 1$ etc. It is also already in
the basis defined in Eq.~(\ref{eq:LagGen}), so for the color-singlet
case only 
\begin{eqnarray}\label{color-singl-1}
&&C^{RR}_{1} =
\left(\frac{2m_p}{G_F^2}\right)\frac{Y_{Qd(1)}^2Y_{e\psi(1)}
  Y_{e^c\psi(1)}}{m^4_{S_{1,2,1/2}}{m_{\psi_{1,X,0}}}} \equiv C^{RR}_{1(0)}
\end{eqnarray}
is non-zero. For the color octet, however, before applying the standard
non-relativistic impulse approximation to convert quark to nucleon
currents, first the color singlet has to be extracted. Using
\begin{eqnarray}\label{lambda-1}
\nonumber
(\lambda^A)^{a}_{b}(\lambda^A)^{c}_{d}=-\frac{2}{3}\delta^a_b\delta^c_d
+ 2 \delta^a_d\delta^c_b, 
\end{eqnarray}
this leads to an operator, which contains
the original operator plus a color mismatched piece:
\begin{equation}\label{eq:step2}
-\frac{2}{3}({\bar Q}^{a}d_{R,a})({\bar Q}^{a}d_{R,a})
+ 2 ({\bar Q}^{a}d_{R,b})({\bar Q}^{b}d_{R,a}).
\end{equation}
This can be brought to canonical form of 
Eqs. (\ref{eq:OperBasis-1})-(\ref{eq:OperBasis-5})
using Fierz rearrangement for the second mismatched term. 
Then we have
\begin{equation}\label{eq:step3} 
{\cal L}_{eff} \propto - C^{RR}_{1}\cdot \mathcal{O}^{RR}_{1}
      - C^{RR}_{2}\cdot \mathcal{O}^{RR}_{2} ,
%
\end{equation}
with
\begin{eqnarray}\label{eq:step3-1}
&&C^{RR}_{1} = \frac{5}{3}C^{RR}_{1(0)}, \ \ \ C^{RR}_{2} = \frac{1}{4}C^{RR}_{1(0)}, \ \ \mbox{and} \ \  \ 
C^{RR}_{2} = \frac{3}{20} C^{RR}_{1}.
\end{eqnarray}
Thus there appear simultaneously two operators the $\mathcal{O}^{RR}_{1}$ and $\mathcal{O}^{RR}_{2}$ at the decoupling scale of the heavy fields $S$ and $\psi$. In the next sections we will call this scale the matching scale $\Lambda$ where we match a high-energy ``full theory'' with its effective low-energy theory.
%
At the end of section~\ref{Sec:Num} we discuss the issues of  
the simultaneous presence of two or more operators at the matching scale.
Note that 
all high-energy scale models specified in Ref.~\cite{Bonnet:2012kh} lead to at most two 
different operators in effective theory at the matching scale.

\section{OPE and  QCD effects}
\label{sec:QCD}

\begin{figure}[t]
\includegraphics[width=0.49\linewidth]{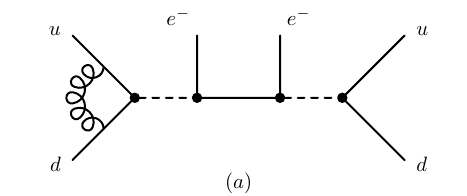}
\includegraphics[width=0.49\linewidth]{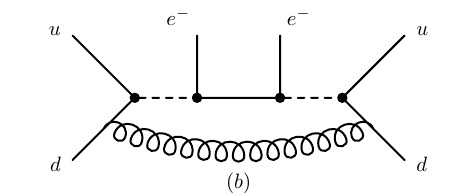}
\caption{$\znbb$-decay example diagrams of one-loop QCD corrections to
  the ``full theory''}
\label{fig:toposQCD}
\end{figure}

\begin{figure}[t]
\includegraphics[width=0.3\linewidth]{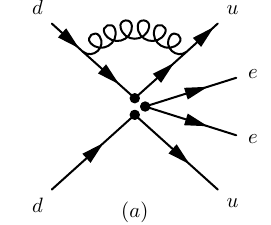}
\includegraphics[width=0.3\textwidth]{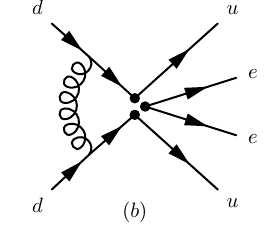}
\includegraphics[width=0.3\textwidth]{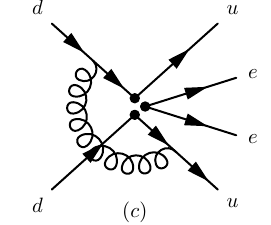}
\caption{One-loop QCD corrections to the short range mechanisms of 
$0\nu\beta\beta$-decay in the effective theory.}
\label{fig:qcdc}
 \end{figure}

In this section we develop the formalism for taking into account the
Leading Order (LO) QCD corrections to the operator product expansion
given in Eq.~(\ref{eq:LagGen}).  We follow essentially the methods
described in the reviews \cite{Buchalla:1995vs, Buras:1998raa} for
semi-leptonic and hadronic decays of mesons.  An important feature of
the 
effective Lagrangian~(\ref{eq:LagGen}), is that it involves the complete set of $d=9$
operators (\ref{eq:OperBasis-1})-(\ref{eq:OperBasis-5}) contributing
to $\znbb$-decay. Therefore, no new operators are generated under
renormalization.

\subsection{Matching of  Full Theory onto Effective One}
\label{subsec:Match}

We start with the discussion of the procedure relating the high-scale
renormalizable model, typically dubbed in the present context as a
``full theory'', to an effective theory represented by the Lagrangian~(\ref{eq:LagGen}) treated as the low-energy limit of the full
theory. This matching procedure allows one to derive the coefficients
$C_{i}$ in terms of the parameters of a high-scale model and take
into account the corresponding perturbative QCD effects.  The matching
is settled at the level of amplitudes of the full theory ${\cal
  A}_{full}$ and of the effective one ${\cal A}_{eff} \sim \langle
{\cal L}^{0\nu\beta\beta}_{\rm eff}\rangle$ requiring they
coincide 
\begin{eqnarray}\label{Match-1}
&&{\cal A}_{full} = {\cal A}_{eff} = \frac{G_F^2}{2 m_p} \sum_{i} C_{i}(\mu)\cdot 
\langle\mathcal{O}_{i}(\mu)\rangle
\end{eqnarray}         
at an energy scale $\mu \leq \Lambda$ below the heavy particle masses
of the full theory.  This is the so-called matching condition.  In the right-hand side of this equation only the leading term $\sim
(1/\Lambda)^{5}$ of the low-energy expansion is retained. The brackets $\langle \rangle$ denote matrix elements between the hadronic states.  Since the
coefficients $C_{i}$ we are interested in, do not depend on the
external states one can use the simplest hadronic states for the
amplitude calculation, which are the quarks. For the same reason we
are allowed to set quark masses to zero and assign to all of them the
common value of the space-like momentum $p^{2}<0$. The latter allows
us to avoid infrared singularities in the calculation.  The diagrams
representing the one-loop QCD corrections to the matrix elements of
the effective operators in Eq.~(\ref{Match-1}) are shown in
Fig~\ref{fig:qcdc}. In Fig.~\ref{fig:toposQCD} we give an example of
the set of one-loop diagrams relevant for the calculation of the full
theory amplitude.  In order to tackle the ultraviolet (UV)
divergencies we use the dimensional regularization and the
$\overline{\rm MS}$ subtraction scheme.  For simplicity we assume that
the masses of all the heavy particles of the full theory are equal to
a common scale $\Lambda =\Lambda_{LNV}$. 
Note, however, that given current LHC constraints, lower limits
on $\Lambda_{LNV}$ are already of the order of $\Lambda_{LNV} \sim
{\cal O}(1)$ TeV.  We will come back to this point in section
\ref{Sec:Num}.  A straightforward calculation shows the general
structure of the amplitude and the operator matrix elements in the LO
of the QCD perturbation theory has the following form:
\begin{eqnarray}
\label{eq:Afull-1}
\mathcal{A}^{Full}&=& 
\frac{g_{\rm full}}{\Lambda^{5}}
a_{i}\left[1+ 
c_{i}\frac{\alpha_{s}}{4\pi}\left(\frac{1}{\epsilon}+\ln\left(\frac{\mu^2}{-p^2}\right)\right)
+\frac{\alpha_{s}}{4\pi}z_{i}\ln\left(\frac{\Lambda^2}{-p^2}\right)
\right]
\langle\mathcal{O}_{i}\rangle_{\rm tree}\\[3mm]   
\label{eq:MatrOper}
\langle\mathcal{O}_{i}\rangle^{(0)}&=& 
\left[\delta_{ij}+ 
\frac{\alpha_{s}}{4\pi}b_{ij}\left(\frac{1}{\epsilon}+\ln\left(\frac{\mu^2}{-p^2}\right)\right)
\right]
\langle\mathcal{O}_{j}\rangle_{\rm tree}.
\end{eqnarray}
Here, $\langle\mathcal{O}_{i}\rangle_{\rm tree}$ are the operator
matrix elements without QCD corrections. The explicit form of the
matrix $b_{ij}$ will be given in the LO approximation below.  On the
other hand, we do not need any knowledge of the coefficients
$a_{i},c_{i}$ or $z_{i}$ since our goal is to calculate the QCD running of
the Wilson coefficients of the effective operators, which is
determined, as discussed below, by $b_{ij}$ only. The above
expression (\ref{eq:Afull-1}) is given in order to clarify some
aspects of the matching. 
The singular $1/\epsilon$ term in Eq.~(\ref{eq:Afull-1}) originates
from the diagram with the vertex correction Fig.~(\ref{fig:toposQCD}a),
which is UV divergent, while the diagram Fig.~(\ref{fig:toposQCD}b) leads
to the finite second term due to the propagators of the virtual heavy
particles of the mass $\sim \Lambda$ cutting the logarithmic divergence
at $\Lambda$. The singularity from the first term can be eliminated by
coupling constant and quark field renormalization. The quark
field renormalization due to the QCD corrections is given by
\begin{eqnarray}\label{eq:QuarkRen}
q^{(0)} = Z_{q}^{1/2} q, \ \ \ \mbox{with} \ \ \ 
Z_{q} = 1 - C_{F} \frac{\alpha_{s}}{4\pi} \, \frac{1}{\epsilon} 
   + O(\alpha_{s}^{2}),
\end{eqnarray}
where $q^{(0)}$ and $q$ are the bare and renormalized quark fields with
the renormalization constant $Z_{q}$ given in the LO approximation.  Here
$C_{F} = (N^{2}-1)/(2N)$ is the standard $SU(N)$ color factor.  In the
case of the operator matrix elements in Eq.~(\ref{eq:MatrOper}) the
$1/\epsilon$-singularities are removed by the quark field
renormalization, Eq.~(\ref{eq:QuarkRen}), accompanied by
renormalization of the operators $\mathcal{O}_{i}^{(0)} =
Z_{ij}\mathcal{O}_{j}$, mixing them within certain groups of the
complete list (\ref{eq:OperBasis-1})-(\ref{eq:OperBasis-5}).  These
groups are identified in the next section.  The operator matrix
elements in Eq.~(\ref{eq:MatrOper}) are renormalized as amputated
Green functions
\begin{eqnarray}\label{eq:MatrOperRen}
\langle\mathcal{O}_{i}\rangle^{(0)} = Z_{q}^{-2} Z_{ij}
                    \langle\mathcal{O}_{j}\rangle .
\end{eqnarray}
Requiring the cancelation of the singularities in Eq.~(\ref{eq:MatrOper})
one finds
\begin{eqnarray}\label{eq:Zij}
Z_{ij} = 
\delta_{ij} + 
\frac{\alpha_{s}}{4\pi} (b_{ij} -  2 C_{F} \delta_{ij}) 
\frac{1}{\epsilon} +
O(\alpha_{s}^{2}) .
\end{eqnarray}
The renormalized matrix elements $\langle\mathcal{O}_{j}\rangle$ and
the amplitude $\mathcal{A}^{Full}$ have the same form as in
Eqs.~(\ref{eq:MatrOper}), (\ref{eq:Afull-1}), but with $1/\epsilon =
0$ and $b_{ij}$ substituted by $b_{ij} - 2 C_{F} \delta_{ij}$.
Inserting these finite quantities in the matching condition, 
Eq.~(\ref{Match-1}), one finds the Wilson coefficients in the form
\begin{eqnarray}
\label{eq:WilsCoeff}
C_{i}(\mu) 
= \left(\delta_{ij}  +  
\frac{\alpha_{s}}{4\pi} f_{ij}\ln\left(\frac{\Lambda^2}{\mu^2}
\right) + O(\alpha_{s}^{2}) \right)C_{j}^{tree},
\end{eqnarray}
where $C_{i}^{tree} = C_{i}(\mu=\Lambda)$ are the coefficients derived from
a high scale model by integrating out heavy particles and neglecting
the QCD corrections. In this formula  $f_{ij}$ are some numerical coefficients which
explicit form is irrelevant for the present discussion.  The above relation 
is shown in order to motivate the subsequent analysis needed to make contact with $0\nu\beta\beta$ scales, $\mu=\mu_{0}\sim 100$ MeV.  As seen Eq. (\ref{eq:WilsCoeff}) in this case contains a large logarithmic term
$\alpha_{s}\ln(\Lambda/\mu_{0})$ 
potentially dangerous for perturbation theory when $\Lambda\sim\Lambda_{LNV}$.
The appearance of  the large logarithms is inevitable issue of renormalizable theories, when one wants to relate the values of a physical quantity like $C_{i}$ measured at two hierarchical scales like $\Lambda_{LNV}$ and $\mu_{0}$. The way out is very well known: one has to    
sum up the large logarithms in all orders in $\alpha_{s}$ on the basis of the Renormalization Group Equations (RGE). It is done in what follows. 


\subsection{QCD running of Wilson coefficients}

In the previous section we recalled a well-known fact that the effective operators may mix under renormalization since the operator renormalization constants $Z_{ij}$  represent, in general,  a non-diagonal matrix.   
It is introduced in Eq.~(\ref{eq:MatrOperRen}) in order to eliminate the divergencies in the operator matrix elements Eq.~(\ref{eq:MatrOper}).
In practice it  is convenient to reformulate the operator mixing in terms of the mixing of the corresponding Wilson coefficients.  These two approaches are well-known to be equivalent~\cite{Buchalla:1995vs, Buras:1998raa}.  It might be helpful to remind this point and the main steps leading to the RGE for the Wilson coefficients.   

Following Refs.~\cite{Buchalla:1995vs, Buras:1998raa} we may consider the effective Lagrangian (\ref{eq:LagGen}) 
within the conventional counterterm approach as a sum of the 4-quark-leg effective vertices 
$\mathcal{O}(q^{(0)})$ constructed of bare quark fields $q^{(0)}$ accompanied with bare ``couplings'' $C^{(0)}$ related to the renormalized ones
\begin{eqnarray}\label{eq:ren-C}
&& q^{(0)} = Z_{q}^{1/2} q, \ \ \ \ C^{(0)}_{i} = Z^{C}_{ij} C_{j}.
\end{eqnarray}
%
Thus for the Lagrangian terms in Eq.~(\ref{eq:LagGen}) one can write 
\begin{eqnarray}\label{eq:renorm-Lag-1}
&& C^{(0)}_{k} \mathcal{O}_{k}(q^{(0)}) = Z^{2}_{q} Z^{(C)}_{ij}C_{j}\mathcal{O}_{i}(q).
\end{eqnarray}
The amplitude calculated with this Lagrangian schematically takes the form
\begin{eqnarray}\label{eq:ren-Ampl-2}
&& \mbox{Ampl.} \sim Z^{2}_{q} Z^{C}_{ij}C_{j}\langle \mathcal{O}(q)_{i}\rangle^{(0)} = C_{j}\langle \mathcal{O}(q)_{i}\rangle .
\end{eqnarray}
The righthand side can be made finite by adjusting the renormalization constants.
In practice it is done order by order in perturbation theory separating the above Lagrangian in the renoramalized part and the counterterms $C_{i}\mathcal{O}(q)_{i} + c.t.$ 
The above result must be consistent with Eq.~(\ref{eq:MatrOperRen}).  Then  $Z^{(C)}_{ij} = Z^{-1}_{ij}$.  Using the fact that the bare quantities $C^{(0)}_{i}$ are independent of the renormalization scale $\mu$ 
\begin{eqnarray}\label{eq:RGE-deriv}
&& \frac{d}{d \ln\mu} C^{(0)}_{i} = \frac{d}{d \ln\mu} Z^{-1}_{ij} C_{j} = 0
\end{eqnarray}
one finds the corresponding RGE for the Wilson coefficients in
the matrix form
\begin{eqnarray}\label{eq:RGE-1}
\frac{d \vec{C}(\mu)}{d \ln\mu} = \hat\gamma^{T}\vec{C}(\mu), 
\end{eqnarray}
where $\vec{C} = (C_{1}, C_{2},...)$ and the anomalous dimension
matrix $\hat\gamma$ is defined as
\begin{eqnarray}\label{eq:AmomDimMatr-1}
&& \hat\gamma = \frac{1}{\hat{Z}} \frac{d}{d \ln\mu} \hat{Z}.
\end{eqnarray}
%
The LO expression in the $\overline{\rm MS}$-scheme is 
~\cite{Buras:1998raa}:
\begin{eqnarray}\label{eq:RGE-2}
\hat\gamma (\alpha_{s}) = - 2 \alpha_{s} \frac{\partial 
              \hat{Z}_{1}(\alpha_{s})}{\partial \alpha_{s}} ,
\end{eqnarray}
where $\hat{Z}_{1}$ is the matrix factor of the singularity
$1/\epsilon$ in Eq.~(\ref{eq:Zij}).  Thus we have in the LO
approximation
\begin{eqnarray}\label{eq:RGE-3}
\gamma_{ij}(\alpha_{s}) = \frac{\alpha_{s}}{4\pi}\gamma_{ij}, 
\ \ \ \mbox{with} \ \ \ 
\gamma_{ij} =  -2 (b_{ij} -  2 C_{F} \delta_{ij}),
\end{eqnarray}
where $\gamma_{ij}$ are the components of the anomalous dimension
matrix $\hat\gamma$.

The solution of Eq.~(\ref{eq:RGE-1}) can be represented in terms of
the $\mu$-evolution matrix
\begin{eqnarray}\label{eq:RGE-Sol}
\vec{C}(\mu) = \hat{U}(\mu, \Lambda)\cdot \vec{C}(\Lambda).
%
\end{eqnarray}
between the low and high energy scales $\mu$ and $\Lambda$, respectively. 
In the LO one finds
\begin{eqnarray}\label{eq:U-matrix}
\hat{U}(\mu, \Lambda) = \hat{V}\, {\rm Diag}\left\{
\left[\frac{\alpha_{s}(\Lambda)}{\alpha_{s}(\mu)}\right]^{\gamma_{i}/(2 \beta_{0})} 
\right\}\, \hat{V}^{-1}.
%
%
\end{eqnarray}
The LO QCD running coupling constant is as usual
\begin{eqnarray}\label{eq:Alpha-Run}
\alpha_{s}(\mu) = \frac{\alpha_s(\Lambda)}{1-\beta_0\frac{\alpha_s(\Lambda)}{2\pi}
\log{\left(\frac{\Lambda}{\mu}\right)}}
%
%
\end{eqnarray}
with $\beta_{0} = (33 - 2 f)/3$, where $f$ is the number of the quark
flavors with masses $m_{f} < \mu$. For a normalization we use  
the experimental value $\alpha_s(\mu = M_z)=0.118$ \cite{Beringer:1900zz}.
Eq.~(\ref{eq:U-matrix}) contains the matrix $\hat{V}$ defined as
\begin{eqnarray}\label{eq:V-matr}
{\rm Diag}\left\{\gamma_{i}\right\} = \hat{V}^{-1} \hat\gamma \hat{V},
\end{eqnarray}
where $\hat\gamma$ is the matrix form of $\gamma_{ij}$, 
see Eq.~(\ref{eq:RGE-3}). The matrix in the left hand side of
Eq.~(\ref{eq:V-matr}) is a diagonal matrix with the diagonal
elements $\gamma_{i}$. The same notation is used in
Eq.~(\ref{eq:U-matrix}).

The quark thresholds in the evolution of the $C_{i}(\mu)$ down to
$\mu_{0}\sim 1$ GeV can be approximately taken into account by the
chain of the $\mu$-evolution matrices with different numbers, $f$, of
quark flavors:
\begin{eqnarray}\label{eq:Thresh}
\hat{U}(\mu, \Lambda = M_{W}) &=& \hat{U}^{(f=3)}(\mu_{0}, \mu_{c})  
\hat{U}^{(f=4)}(\mu_{c}, \mu_{b})  
\hat{U}^{(f=5)}(\mu_{b}, M_{W}),\\ 
\label{eq:Thresh-1}
\hat{U}(\mu, \Lambda> m_{t}\  \, ) &=& \hat{U}^{(f=3)}(\mu_{0}, \mu_{c})  
\hat{U}^{(f=4)}(\mu_{c}, \mu_{b})  
\hat{U}^{(f=5)}(\mu_{b}, \mu_{t}) \hat{U}^{(f=6)}(\mu_{t}, \Lambda),
%
%
\end{eqnarray}
for two cases of the high energy scale $\Lambda$ considered in the present paper. 
Here $\hat{U}^{(f)}$ are the matrix $\hat{U}$ in Eq. (\ref{eq:U-matrix}) calculated for  $f=3,4,5,6$ quark flavors.
The intermediate scales we simply put to the corresponding quark
thresholds $\mu_{c} = m_{c}, \mu_{b} = m_{b}, \mu_{t} = m_{t}$, which is an adequate appropriation
for the LO analysis (for more details see Refs.~\cite{Buchalla:1995vs,Buras:1998raa}).

\subsection{Leading order QCD running of the $\znbb$ operator basis}
\label{subsec:QCD&znbb}

In the leading order, the QCD corrections of the effective operators
of the $\znbb$ basis  (\ref{eq:OperBasis-1})-(\ref{eq:OperBasis-5}) are shown in Fig.~\ref{fig:qcdc}. Of course, other
similar 1-loop diagrams with all other possible gluon links of the
quark legs have to be taken into account additionally.  The diagrams
(a), (b), (c)
contribute to the
operator matrix elements (\ref{eq:MatrOper}) with the following
structures
\begin{eqnarray}
\label{eq:Ma}
{\rm (a)} &\sim&    (\bar{u}\Gamma^i P_{X}d) \ \cdot \ (\bar u\gamma_{\alpha}\gamma_{\beta}\Gamma^j\gamma^{\beta}\gamma^{\alpha}P_{Y} d)\ \cdot \ j_{\cal O} \ \cdot \ C_F\frac{1}{4}\frac{\alpha}{4\pi}\left(\frac{1}{\epsilon}+\log\frac{\mu^2}{-p^2}\right)\\
\label{eq:Mb}
{\rm (b)} &\sim&
    -(\bar{u}\Gamma^i\gamma_{\sigma}\gamma_{\alpha}T^{a} P_{X} d) \ \cdot \ (\bar u\Gamma^j\gamma^{\sigma}\gamma^{\alpha} T^{a} P_{Y}  d)\ \cdot \ j_{\cal O}\ \cdot \ \frac{1}{4}\frac{\alpha}{4\pi}\left(\frac{1}{\epsilon}+\log\frac{\mu^2}{-p^2}\right)
\\
\label{eq:Mc}    
{\rm (c)} &\sim&
(\bar{u}\Gamma^i\gamma_{\sigma}\gamma_{\alpha} T^{a} P_{X}  d) \ \cdot \ (\bar s\gamma^{\alpha}\gamma^{\sigma}\Gamma^j T^{a} P_{Y}  d)\ \cdot \ j_{\cal O} \ \cdot \ \frac{1}{4}\frac{\alpha}{4\pi}\left(\frac{1}{\epsilon}+\log\frac{\mu^2}{-p^2}\right),
\end{eqnarray} 
where $\Gamma^i$ are the Lorentz structures corresponding to the
operators from Eqs. (\ref{eq:OperBasis-1})-(\ref{eq:OperBasis-5}) with
the leptonic currents $j_{\cal O}$, see Eq.~(\ref{eq:Curr}) and $T^a$
being the generators of $SU(N=3)$.  
Using Eq.~(\ref{eq:RGE-3}) we find the LO anomalous dimensions for all the
$\znbb$-operators as
\begin{eqnarray}
\label{Gamma-31-LR}
&&\hat{\gamma}^{XY}_{(31)}=-2\left(
\begin{array}{cc}
-\frac{3}{N}&-6\\
0&6C_F
\end{array}
\right), \ \ \ \ 
\hat{\gamma}_{(12)}^{XX}=-2\left(
\begin{array}{cc}
6 C_F-3&\frac{1}{2N}+\frac{1}{4}\\
-12-\frac{24}{N}&-3-2C_F
\end{array}
\right)\\
\label{Gamma-3-LL-1-LR}
&&\gamma_{(3)}^{XX}=-2\left(\frac{3}{N}-3\right), \ \ \ \  
\gamma_{(5)}^{XY} = -3 \gamma_{(4)}^{XY} =
-12\, C_F,\\
\label{Gamma-45-LL}   
&&\hat{\gamma}_{(45)}^{XX}=-2\left(
\begin{array}{cc}
9-2 C_F&3 i-\frac{6i}{N}\\
i+\frac{2i}{N}&6C_F+1
\end{array}
\right)\, ,
%
%
\end{eqnarray}
where the superscripts $X \neq Y = L, R$  denote the chiralities while the subscripts indicate the operators 
from Eqs.~(\ref{eq:OperBasis-1})-(\ref{eq:OperBasis-5})  mixed  under the renormalization. 
For instance, the first matrix mixes the operators $\mathcal{O}^{XY}_{3}, \mathcal{O}^{XY}_{1}$ with 
$X\neq Y = L, R$, and so on. The anomalous dimensions  in the second row (\ref{Gamma-3-LL-1-LR}) are 
just numbers renormalizing each of the operators $\mathcal{O}^{XX}_{3}$, $\mathcal{O}^{XY}_{5, 4}$
separately without mixing.
Then, using Eqs. (\ref{eq:U-matrix})-(\ref{eq:Thresh}), one can find
the $\mu$-evolution matrix $U(\mu, \Lambda)$ and explicitly relate the
Wilson coefficients at high- and low-energy scales 
$\Lambda$ and $\mu$, respectively.


\section{QCD corrected $\znbb$ half-life and Limits on High-scale models}
\label{Sec:Num}

Now we express 
Eq. (\ref{eq:T12}) in
terms of the high-scale $C_{i}(\Lambda)$ Wilson coefficients using the RGE relations derived in the previous section and obtain the leading order QCD corrected 
$\znbb$-decay half-life  formula, which is
\underline{\it the central result} of the present paper:
\begin{eqnarray}
\label{eq:T12-CMW} 
\Big[ T^{\znbb}_{1/2}\Big]^{-1} &=& 
G_1 
\left| 
\beta_{1}^{XX}\left(C^{LL}_{1}(\Lambda) + C^{RR}_{1}(\Lambda) \right) + 
\beta_{1}^{LR}\left(C^{LR}_{1}(\Lambda)+C^{RL}_{1}(\Lambda)\right) +\right.\\
\nonumber
&&\hspace{4mm}+ \left. \beta_{2}^{XX}\left(C^{LL}_{2}(\Lambda) + C^{RR}_{2}(\Lambda)\right)+ \right. \\
\nonumber
&&\hspace{4mm}+ \left. \beta_{3}^{XX}\left(C^{LL}_{3}(\Lambda) +C^{RR}_{3}(\Lambda)\right) +
\beta_{3}^{LR}\left(C^{LR}_{3}(\Lambda)+C^{RL}_{3}(\Lambda)\right) %
\right|^2 +  \\[3mm]
\nonumber
&+& G_2 
\left|
\beta_{4}^{XX}\left(C^{RR}_{4}(\Lambda) + C^{RR}_{4}(\Lambda)\right)
+ \beta_{4}^{LR} \left(C^{LR}_{4}(\Lambda) + C^{RL}_{4}(\Lambda)\right)
 +\right.\\
\nonumber
&&\hspace{4mm}+ \left. 
\beta_{5}^{XX}\left(C^{RR}_{5}(\Lambda) + C^{RR}_{5}(\Lambda)\right)
+ \beta_{5}^{LR} \left(C^{LR}_{5}(\Lambda) + C^{RL}_{5}(\Lambda)\right)
\right|^2\, ,
\end{eqnarray}
where
\begin{eqnarray}\label{beta-1}
\beta_{1}^{XX} &=& \hspace{5mm} {\cal M}_{1}\ \ \,U_{(12) 11}^{XX} + {\cal M}_{2} U_{(12) 21}^{XX}, \hspace{5mm}    
\beta_{1}^{LR} =  {\cal M}_{3}^{(+)} U_{(31) 12}^{LR} + {\cal M}_{1} U_{(31) 22}^{LR},\\
\label{beta-2}
\beta_{2}^{XX} &=& \hspace{5mm}{\cal M}_{1}\ \ \, U_{(12) 12}^{XX} + 
{\cal M}_{2} U_{(12) 22}^{XX}, \\
\label{beta-3}
\beta_{3}^{XX} &=& \hspace{5mm}{\cal M}_{3}^{(-)} U_{(3)}^{XX}, \hspace{33mm}   
\beta_{3}^{LR} =  {\cal M}_{3}^{(+)} U_{(31) 11}^{LR}\\
\label{beta-4}
\beta_{4}^{XX} &=& -\left|{\cal M}_{4}\right|\  U_{(45) 11}^{XX} + 
\left|{\cal M}_{5}\right| U_{(45) 21}^{XX}, \ \   
\beta_{4}^{LR} = \left|{\cal M}_{4}\right| U_{(4)}^{LR},\\
\label{beta-5} 
\beta_{5}^{XX} &=&-\left|{\cal M}_{4}\right| \ U_{(45) 12}^{XX} + 
\left|{\cal M}_{5}\right| U_{(45) 22}^{XX}, \ \   
\beta_{5}^{LR} = \left|{\cal M}_{5}\right| U_{(5)}^{LR}.
%
%
\end{eqnarray}
From Eqs. (\ref{eq:OperBasis-1}) and (\ref{eq:OperBasis-3}) one sees
that $\mathcal{O}^{XY}_1$ and $\mathcal{O}^{XY}_3$ are symmetric under
the interchange of $X$ and $Y$. Consequently, in
Eq.~(\ref{eq:T12-CMW}) $C_1^{LR}=C_1^{RL}$ and $C_3^{LR}=C_3^{RL}$,
which is equivalent to a factor $2$.  
Coherently with Eqs. (\ref{Gamma-31-LR})-(\ref{Gamma-45-LL}) the subscripts of the evolution matrix $U$ in the parenthesis denote the subscripts of the operators from Eqs.~(\ref{eq:OperBasis-1})-(\ref{eq:OperBasis-5}) mixed under the renormalization, the subscripts without the parenthesis specify the $U$-matrix element.    
Numerical values of these matrix elements are given in Appendix A.
%
%
The nuclear matrix elements
${\cal M}_{i}$ are defined in Ref.~\cite{Pas:2000vn} and can be
calculated in any nuclear structure model. We use their numerical
values as given in Ref.~\cite{Deppisch:2012nb} and display them for
convenience in Table~\ref{ta:NME}.

\setlength{\tabcolsep}{4.5pt}

\begin{table}[h]
\centering
\begin{tabular}{ccccccc}\hline
  $^A$X & ${\cal M}_{1}$ & ${\cal M}_{2}$ & ${\cal M}_{3}^{(+)}$ & 
  ${\cal M}_{3}^{(-)}$ & $|{\cal M}_{4}|$ & $|{\cal M}_{5}|$ \\ \hline
 $^{76}$Ge  & $9.0$ & $-1.6\times 10^{3}$ & $1.3\times 10^{2}$ & 
  $2.1\times 10^{2}$ & $|1.9\times 10^{2}|$ & $|1.9\times 10^{1}|$ \\ 
    $^{136}$Xe & $4.5$ & $-8.5\times 10^{2}$ & $6.9\times 10^{1}$ & 
  $1.1\times 10^{2}$ & $|9.6\times 10^{1}|$ & $|9.3|$ \\ \hline
\end{tabular}
\caption{The numerical values of the nuclear matrix elements ${\cal
    M}_{i}$ taken from  Ref.~\cite{Deppisch:2012nb}.}
\label{ta:NME}
\end{table}

The currently best lower bounds on the $0\nu\beta\beta$-decay half-life 
come from experiments using $^{76}$Ge 
(combined GERDA and Heidelberg--Moscow limits) \cite{Agostini:2013mzu} 
and $^{136}$Xe (combined EXO and KamlandZEN limits) \cite{Gando:2012zm}. 
We use:
\begin{eqnarray}\label{HD-M-Ge}
T_{1/2}^{\znbb}({^{76}Ge})
&\geq& T_{1/2}^{\znbb-exp}({^{76}Ge}) \  = 3.0~ 10^{25}~ \mbox{yrs},\\
\label{EXO-KZEN-Xe} 
T_{1/2}^{\znbb}({^{136}Xe})
&\geq& T_{1/2}^{\znbb-exp}({^{136}Xe}) = 3.4~ 10^{25}~ \mbox{yrs}.
\end{eqnarray}
From these experimental lower bounds we derive upper limits on
$C_{i}(\Lambda_{1,2})$ for two scales  $\Lambda_{1}= M_{W}$ 
and $\Lambda_{2}=\Lambda_{LNV}\sim$1 TeV using Eq.~(\ref{eq:T12-CMW}). 
The choice of $\Lambda_{2}$ is motivated by the facts that the $d=9$ effective
operators  (\ref{eq:OperBasis-1})-(\ref{eq:OperBasis-5}), contributing to the short-range mechanism of $\znbb$, are generated at the mass scale of the heavy particles $\sim \Lambda_{LNV}$, which, considering
the current LHC bounds, are heavier than $\sim$1 TeV. 
The results are
shown in Table~\ref{ta:Cmu}, where we also present, for comparison,
the ``old limits'' on $C_{i}$ (cf. Refs.~\cite{Pas:2000vn,Deppisch:2012nb})  neglecting the QCD running, but
updated with the new half-live limits as given in Eqs (\ref{HD-M-Ge})
and (\ref{EXO-KZEN-Xe}).  Note that neglecting the QCD running corresponds to setting
$U_{ij}^{XY} = \delta_{ij}$ in Eqs. (\ref{beta-1})-(\ref{beta-5}).

Deriving individual limits on $C_{i}$  in Table~\ref{ta:Cmu} we assumed for simplicity that there are no significant cancellations
between the terms in the right-hand side of Eq. (\ref{eq:T12-CMW}). This is equivalent to assuming the dominance of only one $C_{i}$ at a time. 
Comparing different numbers in Table~\ref{ta:Cmu}, one sees that the running between
$M_W$ and $\mu_{0} \simeq 1$ GeV is more important than the running
between $1$ TeV and $M_W$, but the latter is not negligible.  As can
also be seen from Table~\ref{ta:Cmu}, the QCD RGE running has the
largest impact on the contributions to the operators ${\cal
  O}_{1}^{XY}$ and ${\cal O}_{5}^{XX}$. This can be understood since
in the RGE running they mix with the operators ${\cal O}_{2}^{XY}$ and ${\cal
  O}_{4}^{XX}$, respectively, which have significantly larger nuclear
matrix elements, as seen from Table~\ref{ta:NME}.

\begin{table}[h]
\centering
 \begin{tabular}{|c|cc|c||cc|c|}
 \hline
&\multicolumn{2}{c|}{With QCD}&Without QCD&\multicolumn{2}{c|}{With QCD}&Without QCD\\
\hline
$^A$X & $|C_{1}^{XX}(\Lambda_1)|$ & $|C_{1}^{XX}(\Lambda_2)|$ & $|C_{1}^{XX}|$ & $|C_{1}^{LR,RL}(\Lambda_1)|$ & $|C_{1}^{LR,RL}(\Lambda_2)|$ & $|C_{1}^{LR,RL}|$    
\\
\hline
$^{76}$Ge & $5.0\times 10^{-10}$& $3.8\times 10^{-10}$ &$\bf{ 2.6\times 10^{-7}}$&$1.5\times 10^{-8}$& $9.1\times 10^{-9}$ &${\bf 2.6\times 10^{-7}}$ \\ 
$^{136}$Xe & $3.4\times 10^{-10}$ & $2.6\times 10^{-10}$ & $\bf{ 1.8\times 10^{-7}}$& $9.7\times 10^{-9}$& $6.1\times 10^{-9}$ &${\bf 1.8\times 10^{-7}}$  \\
	\hline
	\hline
$^A$X & $|C_{2}^{XX}(\Lambda_1)| $ & $|C_{2}^{XX}(\Lambda_2)| $ & $|C_{2}^{XX}| $ &$-$ & $-$ &$-$
	\\
	\hline
	$^{76}$Ge & $3.5\times 10^{-9}$ & $5.2\times 10^{-9}$ & $1.4\times 10^{-9}$ &$-$ & $-$ &$-$ \\ 
  $^{136}$Xe & $2.4\times 10^{-9}$& $3.5\times 10^{-9}$ & $9.4\times 10^{-10}$ & $-$& $-$ &$-$ \\
	\hline
	\hline
$^A$X & $|C_{3}^{XX}(\Lambda_1)|$ &$|C_{3}^{XX}(\Lambda_2)|$&$|C_{3}^{XX}|$ & $|C_{3}^{LR,RL}(\Lambda_1)|$ &$|C_{3}^{LR,RL}(\Lambda_2)|$ & $|C_{3}^{LR,RL}|$    
\\
\hline
$^{76}$Ge &$1.5\times 10^{-8}$ & $1.6\times 10^{-8}$ &$1.1\times 10^{-8}$ & $2.0\times 10^{-8}$& $2.1\times 10^{-8}$ &$1.8\times 10^{-8}$\\ 
$^{136}$Xe &$9.7\times 10^{-9}$& $1.1\times 10^{-8}$ & $7.4\times 10^{-9}$ &$1.4\times 10^{-8}$& $1.4\times 10^{-8}$ & $1.2\times 10^{-8}$ \\
\hline
\hline
$^A$X & $|C_{4}^{XX}(\Lambda_1)|$ & $|C_{4}^{XX}(\Lambda_2)|$ & $|C_{4}^{XX(0)}|$ & $|C_{4}^{LR,RL}(\Lambda_1)|$ &$|C_{4}^{LR,RL}(\Lambda_2)|$ & $|C_{4}^{LR,RL(0)}|$     
\\
\hline
$^{76}$Ge &$5.0\times 10^{-9}$&$3.9\times 10^{-9}$&${\bf 1.2\times 10^{-8}}$&$1.7\times 10^{-8}$&$1.9\times 10^{-8}$ & $1.2\times 10^{-8}$\\ 
$^{136}$Xe &$3.4\times 10^{-9}$& $2.7\times 10^{-9}$ &${\bf 7.9\times 10^{-9}}$&$1.2\times 10^{-8}$& $1.3\times 10^{-8}$ &$7.9\times 10^{-9}$ \\
\hline
\hline
$^A$X & $|C_{5}^{XX}(\Lambda_1)|$ & $|C_{5}^{XX}(\Lambda_2)|$ & $|C_{5}^{XX}|$ & $|C_{5}^{LR,RL}(\Lambda_1)|$ & $|C_{5}^{LR,RL}(\Lambda_2)|$ & $|C_{5}^{LR,RL}|$
\\
\hline    
$^{76}$Ge & $2.3\times 10^{-8}$ & $1.4\times 10^{-8}$ & ${\bf 1.2\times 10^{-7}}$ & $3.9\times 10^{-8}$ & $2.8\times 10^{-8}$ & ${\bf 1.2\times 10^{-7}}$\\ 
$^{136}$Xe &$1.6\times 10^{-8}$ & $9.5\times 10^{-9}$ & ${\bf 8.2\times 10^{-8}}$ & $2.8\times 10^{-8}$ & $2.0\times 10^{-8}$ & ${\bf 8.2\times 10^{-8}}$ \\    
\hline
\end{tabular}
\caption{Individual upper limits on the Wilson coefficients $C_{i}(\Lambda)$ in
Eq.~(\ref{eq:T12-CMW}) calculated for two matching scales,   $\Lambda_{1} = M_W$ and  
  $\Lambda_{2}=\Lambda_{LNV} = 1$ TeV, using the experimental bounds (\ref{HD-M-Ge}), (\ref{EXO-KZEN-Xe}). Here $X=L,R$.
  For comparison we also give limits on the $C_{i}$, without the QCD running.   
  We highlighted with boldface text those positions where the QCD running leads to about 3 orders (for~$C^{XX}_{1}$), one order (for $C^{LR,RL}_{1}$) of magnitude  and a factor 3--4 (for $C^{XX}_{4}, C^{LR,RL}_{5}$) effect.}
\label{ta:Cmu}
\end{table}


If one is interested in constraining a particular high-scale model
one should  directly use Eq.~(\ref{eq:T12-CMW}) retaining only those coefficients $C_{i}$, which are present in the model. The corresponding limits on the model
parameters in certain cases can be significantly modified with respect to their values given in
Table~\ref{ta:Cmu}, which, as we mentioned above,  are based on the hypothesis about one coefficient dominance at a time. In the case of a particular model this assumption may not be valid.  
For example, the model specified in
Eqs.~(\ref{eq:LagExaEff}), (\ref{eq:step3}) 
contains simultaneously two non-vanishing
Wilson coefficients at the matching scale, $C_{1}^{RR}(\Lambda)$ and
$C_{2}^{RR}(\Lambda)$, which obey the relation $C_{2}^{RR} = (3/20)
C_{1}^{RR}$. 
Using Eqs. (\ref{eq:T12-CMW})-(\ref{beta-5}) one finds for $C^{RR}_{1}(\Lambda)$ nearly the same limit
as in  Table~\ref{ta:Cmu}, but due to the above mentioned relation the limit on $C_{2}(\Lambda)$ turns out to be 
$C^{RR}_{2}(\Lambda_{2}) = (3/20)(C^{RR}_{1}<3.8\times 10^{-10}) < 5.7\times 10^{-11}$, which is two orders of magnitude stronger than the individual limit  $C^{RR}_{2}(\Lambda_{2}) < 5.2\times 10^{-9}$ given in  Table~\ref{ta:Cmu} for the case of 
${}^{76}$Ge. 
%
%
For all models listed in \cite{Bonnet:2012kh} one
can find QCD improved limits in the same manner.  

\section{Conclusions}
\label{Sec:Cncl}

In this paper we have calculated QCD corrections to the complete list
of Lorentz-invariant operators for the short-range (SR) part of the $\znbb$-decay amplitude. We have used the RGE technique to derive 1-loop improved
limits on all the Wilson coefficients appearing in the SR contributions to
$\znbb$-decay.  We stress again, that we have taken special care to
present our results in such a way, that improved limits can be derived
easily, should updated experimental limits or improved nuclear physics
calculations become available.

Our numerical results show that the QCD corrections are indeed important.
We note that both more and less stringent limits can result from
taking into account QCD corrections, depending on the operator under
consideration.  In particular, the appearance of color mismatched
operators lead to operator mixing which, due to largely different
nuclear matrix elements for different operators, can lead to
surprisingly large changes in some limits. QCD improved limits from
$\znbb$-decay should therefore be used, when comparing constraints
from $\znbb$-decay with those derived from LHC.

\bigskip

\centerline{\bf Acknowledgements}

\medskip

M.G. thanks the IFIC for hospitality during her stay.  This work was
supported by the Spanish MICINN grants FPA2014-58183-P and Multidark
CSD2009-00064 (MINECO), and PROMETEOII/2014/084 (Generalitat
Valenciana), and by Fondecyt (Chile) under grants 1150792 and 3160642.


%
%
%
\setcounter{section}{0}
\def\theequation{\Alph{section}.\arabic{equation}}
\setcounter{equation}{0}

\section{Appendix A. Explicit form of the RGE evolution Matrix.}
\label{sec:Appendix-A}
Here we give the numeric values of the RGE $\mu$-evolution matrix elements defined in Eq.~(\ref{eq:U-matrix}) and taking into account the quark thresholds according to (\ref{eq:Thresh}), (\ref{eq:Thresh-1}).  In the notations used in 
Eqs.~(\ref{beta-1})-(\ref{beta-5}) we have for two reference values, $\Lambda_1=M_{W}$ and 
\mbox{$\Lambda_{2}=1$ TeV,} of the high energy scale $\Lambda$, and $\mu_0=1$ Gev, the following results
\begin{eqnarray}
\label{eq:U-num-1}
\hat{U}^{XX}_{(12)}(\mu_0, \Lambda_1)=
\left(
\begin{array}{cc}
  1.88&0.06\\
  -2.76& 0.40
\end{array}
\right),&&U^{XX}_{(3)}(\mu_0, \Lambda_1)=0.76,\\
\label{eq:U-num-2}
\hat{U}^{LR}_{(31)}(\mu_0, \Lambda_1)=
\left(
\begin{array}{cc}
  0.87&-1.40\\
  0& 2.97
\end{array}
\right),
&&
\hat{U}^{XX}_{(45)}(\mu_0, \Lambda_1)=
\left(
\begin{array}{cc}
  2.33&0.39 i\\
  0.64 i& 3.35
\end{array}
\right),\  \ \ \ \ \\
\label{eq:U-num-3}
U^{LR}_{(4)}\, (\mu_0, \Lambda_1)=0.70,\hspace{20mm} \ \, &&
U^{LR}_{(5)}\, (\mu_0, \Lambda_1)=2.97\, .
\end{eqnarray} 
and 
\begin{eqnarray}
\label{eq:U-num-1-1}
\hat{U}^{XX}_{(12)}(\mu_0, \Lambda_{2})=
\left(
\begin{array}{cc}
  2.24&0.07\\
  -3.70& 0.27
\end{array}
\right),&&U^{XX}_{(3)}(\mu_0, \Lambda_{2})=0.70,\\
\label{eq:U-num-2-1}
\hat{U}^{LR}_{(31)}(\mu_0, \Lambda_{2})=
\left(
\begin{array}{cc}
  0.84&-2.19\\
  0& 4.13
\end{array}
\right),
&&
\hat{U}^{XX}_{(45)}(\mu_0, \Lambda_{2})=
\left(
\begin{array}{cc}
  2.98&0.69 i\\
  1.15i& 4.82
\end{array}
\right),\  \ \ \ \ \\
\label{eq:U-num-3-1}
U^{LR}_{(4)}\, (\mu_0, \Lambda_{2})=0.62,\hspace{20mm}\ \,   &&
U^{LR}_{(5)}\, (\mu_0, \Lambda_{2})=4.13\, .
\end{eqnarray} 
Using these RGE evolution matrix elements one can calculate the corresponding 
\mbox{$\beta^{XY}_{i}$-coefficients} (\ref{beta-1})-(\ref{beta-5}) for values of the nuclear matrix elements ${\cal M}_{i}$ 
other than we give in Table~\ref{ta:NME} and used for the derivation of the limits presented in Table~\ref{ta:Cmu}.



\newpage
\begin{center}
\begin{large}\textbf{
Erratum: QCD running in neutrinoless double beta decay:
  Short-range mechanisms}
\end{large}
\end{center}

In the summation of the one-loop QCD corrections to the short-range
mechanism of neutrinoless double beta decay, analyzed in
Ref.~\cite{Gonzalez:2015ady}, we lost a minus sign in the case of the
anomalous dimensions of some effective short-range operators.

Then two of our anomalous dimension matrices have to be modified. For
this reason the upper limits on the $C_1^{XX}$, $C_2^{XX}$, $C_4^{XX}$
and $C_5^{XX}$ Wilson coefficients have to be updated. This erratum
doesn't modify the main conclusions of this work.

\begin{itemize}

\item Equation (42) has to be replaced by 

\begin{eqnarray}
\label{Gamma-31-LR}
&&\hat{\gamma}^{XY}_{(31)}=-2\left(
\begin{array}{cc}
-\frac{3}{N}&-6\\
0&6C_F
\end{array}
\right), \ \ \ \ 
\hat{\gamma}_{(12)}^{XX}=-2\left(
\begin{array}{cc}
6 C_F-3&-\frac{1}{2N}+\frac{1}{4}\\
-12-\frac{24}{N}&-3-2C_F
\end{array}
\right)
\end{eqnarray}

\textbf{Comment:} In reference \cite{Buras:2000if}, the anomalous
dimension matrix $\gamma_{(12)}^{XX}$ was also calculated. We agree
with their results, as can be seen after inserting the definition of
$C_F$. However, in this paper a different convention for the
$\sigma_{\mu\nu}$ matrix was used. While in \cite{Buras:2000if} they
use $\sigma_{\mu\nu}=\frac{1}{2}[\gamma_{\mu},\gamma_{\nu}]$, in our
paper we use
${\sigma_{\mu\nu}}=\frac{i}{2}[\gamma_{\mu},\gamma_{\nu}]$. This is
the reason why in \cite{Buras:2000if} there is a extra minus sign in
the off-diagonal elements of $\gamma_{(12)}^{XX}$.

\item Equation (44) has to be replaced by
\begin{eqnarray}
\label{Gamma-45-LL}   
&&\hat{\gamma}_{(45)}^{XX}=-2\left(
\begin{array}{cc}
-3-2 C_F&-3 i-\frac{6i}{N}\\
-i+\frac{2i}{N}&6C_F-3
\end{array}
\right)\, ,
\end{eqnarray}

\item The upper limits on Wilson coefficients $C_1^{XX}$, $C_2^{XX}$, $C_4^{XX}$ and $C_5^{XX}$ have to be updated. The updated Table II is the following:

\begin{table}[h]
\centering
 \begin{tabular}{|c|ccc|ccc|}
\hline
$^A$X & $|C_{1}^{XX}(\Lambda_1)|$ & $|C_{1}^{XX}(\Lambda_2)|$ & $|C_{1}^{XX(0)}|$ & $|C_{1}^{LR,RL}(\Lambda_1)|$ & $|C_{1}^{LR,RL}(\Lambda_2)|$ & $|C_{1}^{LR,RL(0)}|$    
\\
\hline
$^{76}$Ge & $4.9\times 10^{-10}$& $3.6\times 10^{-10}$ &$\underline{\bf 2.6\times 10^{-7}}$&$1.5\times 10^{-8}$& $9.1\times 10^{-9}$ &${\bf 2.6\times 10^{-7}}$ \\ 
$^{136}$Xe & $3.3\times 10^{-10}$ & $2.5\times 10^{-10}$ & $\underline{\bf 1.8\times 10^{-7}}$& $9.7\times 10^{-9}$& $6.1\times 10^{-9}$ &${\bf 1.8\times 10^{-7}}$  \\
	\hline
	\hline
$^A$X & $|C_{2}^{XX}(\Lambda_1)| $ & $|C_{2}^{XX}(\Lambda_2)| $ & $|C_{2}^{XX(0)}| $ & & $-$ &
	\\
	\hline
	$^{76}$Ge & $3.1\times 10^{-9}$ & $4.0\times 10^{-9}$ & $1.4\times 10^{-9}$ & & $-$ & \\ 
  $^{136}$Xe & $2.1\times 10^{-9}$& $2.7\times 10^{-9}$ & $9.4\times 10^{-10}$ & & $-$ & \\
	\hline
	\hline
$^A$X & $|C_{3}^{XX}(\Lambda_1)|$ &$|C_{3}^{XX}(\Lambda_2)|$&$|C_{3}^{XX(0)}|$ & $|C_{3}^{LR,RL}(\Lambda_1)|$ &$|C_{3}^{LR,RL}(\Lambda_2)|$ & $|C_{3}^{LR,RL(0)}|$    
\\
\hline
$^{76}$Ge &$1.5\times 10^{-8}$ & $1.6\times 10^{-8}$ &$1.1\times 10^{-8}$ & $2.0\times 10^{-8}$& $2.1\times 10^{-8}$ &$1.8\times 10^{-8}$\\ 
$^{136}$Xe &$9.7\times 10^{-9}$& $1.1\times 10^{-8}$ & $7.4\times 10^{-9}$ &$1.4\times 10^{-8}$& $1.4\times 10^{-8}$ & $1.2\times 10^{-8}$ \\
\hline
\hline
$^A$X & $|C_{4}^{XX}(\Lambda_1)|$ & $|C_{4}^{XX}(\Lambda_2)|$ & $|C_{4}^{XX(0)}|$ & $|C_{4}^{LR,RL}(\Lambda_1)|$ &$|C_{4}^{LR,RL}(\Lambda_2)|$ & $|C_{4}^{LR,RL(0)}|$     
\\
\hline
$^{76}$Ge &$2.6\times 10^{-8}$&$3.4\times 10^{-8}$&${\it 1.2\times 10^{-8}}$&$1.7\times 10^{-8}$&$1.9\times 10^{-8}$ & $1.2\times 10^{-8}$\\ 
$^{136}$Xe &$1.8\times 10^{-8}$& $2.3\times 10^{-8}$ &${\it 7.9\times 10^{-9}}$&$1.2\times 10^{-8}$& $1.3\times 10^{-8}$ &$7.9\times 10^{-9}$ \\
\hline
\hline
$^A$X & $|C_{5}^{XX}(\Lambda_1)|$ & $|C_{5}^{XX}(\Lambda_2)|$ & $|C_{5}^{XX(0)}|$ & $|C_{5}^{LR,RL}(\Lambda_1)|$ & $|C_{5}^{LR,RL}(\Lambda_2)|$ & $|C_{5}^{LR,RL(0)}|$
\\
\hline    
$^{76}$Ge & $1.6\times 10^{-8}$ & $1.2\times 10^{-8}$ & ${\bf 1.2\times 10^{-7}}$ & $3.9\times 10^{-8}$ & $2.8\times 10^{-8}$ & ${\it 1.2\times 10^{-7}}$\\ 
$^{136}$Xe &$1.1\times 10^{-8}$ & $8.1\times 10^{-9}$ & ${\bf 8.2\times 10^{-8}}$ & $2.8\times 10^{-8}$ & $2.0\times 10^{-8}$ & ${\it 8.2\times 10^{-8}}$ \\    
\hline
\end{tabular}
\label{ta:Cmu}
\end{table}

\item In the Appendix, Equation (A1) has to be updated by

\begin{eqnarray}
\label{eq:U-num-1}
\hat{U}^{XX}_{(12)}(\mu_0, \Lambda_1)=
\left(
\begin{array}{cc}
  1.95&0.01\\
  -2.82& 0.45
\end{array}
\right),&&U^{XX}_{(3)}(\mu_0, \Lambda_1)=0.76,
\end{eqnarray}

\item In the Appendix, Equation (A2) has to be updated by

\begin{eqnarray}
\label{eq:U-num-2}
\hat{U}^{LR}_{(31)}(\mu_0, \Lambda_1)=
\left(
\begin{array}{cc}
  0.87&-1.40\\
  0& 2.97
\end{array}
\right),
&&
\hat{U}^{XX}_{(45)}(\mu_0, \Lambda_1)=
\left(
\begin{array}{cc}
  0.45&-0.70 i\\
  -0.05 i&1.95
\end{array}
\right),
\end{eqnarray}

\item In the Appendix, Equation (A4) has to be updated by

\begin{eqnarray}
\label{eq:U-num-1-1}
\hat{U}^{XX}_{(12)}(\mu_0, \Lambda_{2})=
\left(
\begin{array}{cc}
  2.39&0.02\\
  -3.83& 0.35
\end{array}
\right),&&U^{XX}_{(3)}(\mu_0, \Lambda_{2})=0.70,
\end{eqnarray}

\item In the Appendix, Equation (A5) has to be updated by

\begin{eqnarray}
\label{eq:U-num-2-1}
\hat{U}^{LR}_{(31)}(\mu_0, \Lambda_{2})=
\left(
\begin{array}{cc}
  0.84&-2.19\\
  0& 4.13
\end{array}
\right),
&&
\hat{U}^{XX}_{(45)}(\mu_0, \Lambda_{2})=
\left(
\begin{array}{cc}
  0.35& -0.96 i\\
  -0.06 i& 2.39
\end{array}
\right)
\end{eqnarray} 

\end{itemize}

\end{document}